# Talbot-like pattern evolution in complex structured light from unitary transformation

Zheng-Xiao Cao[1,2], Ting-Ting Liu[2], Bo-Zhao[2], Carmelo Rosales-Guzmán[2,3], Jun Liu[1*], and Zhi-Han Zhu[2*]

[1] International Collaborative Laboratory of 2D Materials for Optoelectronics Science and Technology of Ministry of Education, Institute of Microscale Optoelectronics, Shenzhen University, Shenzhen 518060, China
[2] Wang Da-Heng Cente of Quantum Control, Harbin University of Science and Technology, Harbin 150080, China
[3] Centro de Investigaciones en Óptica, A.C., Loma del Bosque 115, Colonia Lomas del Campestre, 37150 León, Gto., Mexico
\* e-mail: liuj1987@szu.edu.cn and zhuzhihan@hrbust.edu.cn

Astigmatic unitary transformations allow for the adiabatic connections of all feasible states of paraxial Gaussian beams on the same modal sphere, i.e., Hermite-Laguerre-Gaussian (HLG) modes. Here, we present a comprehensive investigation into the unitary modal evolution of complex structured Gaussian beams, comprised by HLG modes from disparate modal spheres, via astigmatic transformation. The non-synchronized higher-order geometric phases in cyclic transformations originates a Talbot-effect-like modal evolution in the superposition state of these HLG modes, resulting in pattern variations and revivals in transformations with specific geodesic loops. Using Ince-Gaussian modes as an illustrative example, we systematically analyze and experimentally corroborate the beamforming mechanism behind the pattern evolution. Our results outline a generic modal conversion theory of structured Gaussian beams via astigmatic unitary transformation, offering a new approach for shaping spatial modal structure. These findings may inspire a wide variety of applications based on structured light.

## I. Introduction

Light beams exhibiting propagation-invariant transverse patterns (or spatial-complex amplitudes) are eigen modes (or solutions) of the paraxial wave equation (PWE). Notably, the Laguerre-Gaussian (LG) and the Hermite-Gaussian (HG) mode families are the most prevalent examples in cylindrical and Cartesian coordinates, respectively [1–3]. Despite both families sharing the TEM$_{00}$ fundamental mode, they possess distinct features in their transverse patterns and can individually form a complete and unbounded 2D Hilbert space [4,5]. For this reason, structured Gaussian mode families and associated orbit angular momentum (OAM) degrees of freedom offer a promising photonic platform for experimental quantum/classical information, which enables the encoding and manipulation of high-dimensional states/channels within paraxial beams [6–12]. Additionally, by utilizing HG and LG modes as building blocks, one can design and construct customized light beams with desired spatial structures in terms of amplitude, phase, polarization, and temporal frequency. This approach has recently yielded a series of novel diffractive phenomena of structured light, such as OAM high-dimensional states exhibiting propagation-invariant responses to quantum control and optical Hopfions revealed from spin topological textures of 3D vector beams [13–16]. The interpretation of these structured light's diffractive evolution, including pattern variations and revivals during propagation, can be explained by considering the intermodal phase dynamics between LG or HG modal components mediated by Gouy phase [17,18]. By considering the geometric-phase origin of Gouy phase into account [19,20], the perspective offers a novel understanding for shaping and controlling structured light. Specifically, it raises the question of whether more general unitary transformations (beyond diffraction) and the associated geometric phase can be utilized to shape and control modal structured light.

The answer is definitely yes and, particularly, the astigmatic transformation emerges as the foremost consideration for the control of structured Gaussian modes. This adiabatic control enables the connection of all possible states on the same modal sphere, namely the generalized Hermite-Laguerre-Gaussian (HLG) modes [4]. These states possess the identical modal order $N$ and exhibit patterns that remain invariant under cyclic transformations on the sphere, owing to their underlying SU(2) modal structure [5]. Nevertheless, this fundamental principle ceases to hold if the state undergoing the transformation comprises a superposition of HLG modes from disparate modal spheres even in the case of having the same order. It is anticipated that these complex structured Gaussian beams, when subjected to the astigmatic unitary evolution, will inevitably undergo pattern variations and revivals, thereby introducing novel modal structures and intriguing phenomena. In this study, using Ince-Gaussian (IG) modes as a specific illustration, we offer a comprehensive framework for the generic modal evolution of structured Gaussian beams through astigmatic unitary transformation. The modes in question are eigen solutions of the PWE in elliptic coordinates and are essentially superposition states comprised of HG or LG modes on distinct HLG modal spheres [21]. All the HLG components have the identical order $N$, so that the superposition state can be regarded as a 'mode-locked' pattern with propagation invariance [22]. In the following sections of the paper, the pattern variations and revivals of this 'mode-locked' pattern in astigmatic transformation, i.e., a unitary evolution with spatial-mode dispersion, are systematically studied.

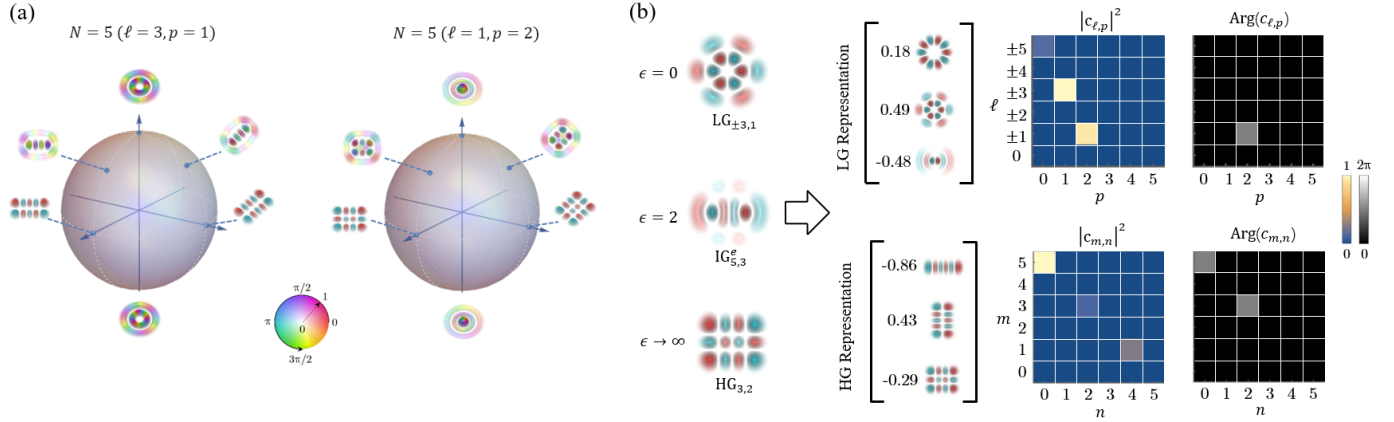

**Figure 1**. (a) Two examples of HLG modal spheres with $N$=5, and (b) the generalization in elliptical coordinates, i.e., Ince-Gaussian modes.

## II. Theoretical Background

The transverse spatial wavefunctions of Laguerre- and Hermite-Gaussian modal families at the waist plane ($z=0$) are

$$\mathrm{LG}_{\ell,p}(\boldsymbol{r}) = \sqrt{\frac{2p!}{\pi(p+|\ell|)!}}\frac{1}{w}\mathrm{e}^{-\frac{r^2}{w^2}}\left(\frac{\sqrt{2}r}{w}\right)^{|\ell|}\mathrm{e}^{-i\ell\varphi}L_p^{|\ell|}\left(\frac{2r^2}{w^2}\right) \quad (1)$$

and

$$\mathrm{HG}_{m,n}(\boldsymbol{r}) = \sqrt{\frac{2^{1-n-m}}{\pi n!\, m!}}\frac{1}{w}\mathrm{e}^{-\frac{r^2}{w^2}}H_n\left(\frac{\sqrt{2}x}{w}\right)H_m\left(\frac{\sqrt{2}y}{w}\right) \quad (2)$$

respectively, where $L_p^{|\ell|}(\cdot)$ and $H_{m/n}(\cdot)$ denote Laguerre and Hermite polynomials defining the two-dimensional modal structure, and $w$ is the beam waist. Further, indices $\ell$ (an integer) and $p$ (a positive integer) determine azimuthal (or OAM) and radial modal structures of LG modes, respectively; while positive integer $m$ and $n$ decide phase-reversal number along coordinate axis of HG modes. Furthermore, the utilization of Eqs. (1) and (2) as the source of the diffraction integral allows for the derivation of three-dimensional wavefunctions, i.e., $\mathrm{LG}_{\ell,p}(\boldsymbol{r},z)$ and $\mathrm{HG}_{m,n}(\boldsymbol{r},z)$, exhibiting an invariant pattern during propagation, which increases in size and acquires an overall Gouy phase $(N+1)\arctan(z/z_R)$, where $z_R$ denotes the Rayleigh range. Both sets of wavefunctions are considered bivariate complete orthonormal sets, meaning that the projection between different bases is equal to $\delta_{\ell,p}$ or $\delta_{m,n}$. Consequently, these sets can be effectively utilized as fundamental components for constructing paraxial light fields with desired spatial structure in terms of amplitude, phase, and polarization, in which the Gouy phase depending on $N$ plays a crucial role for beam structure propagation invariance, variations and revivals [17].

In contrast to the unitary order polarization states, a given HLG modal order $N$, defined by $N = 2p+|\ell| = m+n$, has a total $N+1$ LG and HG modes [23]. Apart from the common fundamental mode $\mathrm{TEM}_{00}$ ($N=0$), HLG modes in each order can form $(N+1)/2$ distinct Poincaré-like modal spheres [24]. These spheres are characterized by $\mathrm{LG}_{\pm\ell,p}$ poles, rotating $\mathrm{HG}_{m,n}$ modes (following the relation $\ell = m-n$) on the equator, and intermediate states defined by the SU(2) modal structure of HLG modes. For instance, Fig. 1(a) illustrates two such spheres with order $N=5$. The experimental connection of all possible states on each sphere can be achieved through astigmatic transformation, which, as the unitary evolution, inherently induces a higher-order geometric phase depending on OAM. [25,26]. In particular, during the cyclic transformation, all possible states (HLG modes) remain in their patterns but acquire an overall geometric phase shift. The magnitude of this phase shift is determined by both the geodesic path loop on the sphere and the value of $|\ell|$ defined by LG poles, but is independent of $N$.

However, this law of pattern invariance in the cyclic transformation is no longer applicable to a complex modal structure consisting of HLG modes with different $|\ell|$, which is precisely the focus of the present paper. To address this issue, we employ PWE eigen solutions in elliptical coordinates known as IG modes as a representative example. This comprehensive Gaussian mode family offers a unified definition for both HG and LG families through the ellipticity parameter $\epsilon$, and its wavefunction with even (odd) parity at the waist plane can be expressed as [27]

$$\mathrm{IG}_{N,m}^{e(o)}(\boldsymbol{r},\epsilon) = C_{e(o)}\mathrm{e}^{-\frac{r^2}{w^2}}S_{N,m}^{e(o)}(i\xi,\epsilon)S_{N,m}^{e(o)}(\eta,\epsilon) \quad (3)$$

where $\xi \in [0,\infty)$ and $\eta \in [0,2\pi)$ denote the radial and angular elliptic variables respectively, $S_{N,m}^{e(o)}$ represents even (odd) Ince polynomials accompanied by the normalization constant $C_{e(o)}$. The parameter $\epsilon$, taking values in the interval $[0,\infty)$, plays a crucial role in determining the ellipticity of the coordinates. This parameter is considered to be super-complete in relation to the two-dimensional transverse plane. Consequently, both the elliptical coordinates and the IG modes defined within them exhibit a lack of uniqueness. More specifically, an IG mode can transform into LG and HG modes with same the order as $\epsilon$ equals 0 and approaches infinity, respectively. This is schematically shown in Fig. 1(b) left column, where the $IG_{5,3}$ mode transitions into the $LG_{\pm 3,1}$ and $HG_{3,2}$ as $\epsilon$ equals 0 and infinity, respectively. In more general scenarios where $\epsilon$ is greater than 0 but significantly less than infinity, the IG mode becomes indeed a superposition state consisting of $(N+1)$ LG or HG modes with the same order $N$ [27]

$$IG_{N,m}^{e(o)}(\boldsymbol{r}, \epsilon) = \sum c_{\ell,p} LG_{\ell,p}(\boldsymbol{r}) = \sum c_{m,n} HG_{m,n}(\boldsymbol{r}) \quad (4)$$

where $c_{\ell,p}$ and $c_{m,n}$ are complex coefficients of the series. Taking the even $IG_{5,3}$ mode with $\epsilon = 2$ as an example, the density matrices depicted on the right hand of Fig. 1(b) indicate that both the LG and HG components possess the same modal order $N = 2p + |\ell| = m + n = 5$. This finding provides an explanation for the propagation invariance of IG patterns from an alternative perspective: all modal components acquire synchronous Gouy phase during propagation, resulting in constant intramodal phase, i.e., constant $\text{Arg}(c_{\ell,p})$ or $\text{Arg}(c_{m,n})$. However, subsequent analysis reveals that these complex structured Gaussian modes, composed of HLG modes with the same order, undergo intricate pattern evolution and revivals in the astigmatic unitary transformation, unveiling a sequence of PWE eigen modes with exotic patterns.

## III. Results and Discussion

The essence of unitary transformation lies in the intramodal phase variation of a quantum (or classical) state, and thus its physical implementation relies on the manipulation of phase retardance between the intramodal components. Phase-only devices with modal dependence therefore serve as effective means for the transformation. For instance, the well-known half-wave (HW) and quarter-wave (QW) plates for polarization control introduce $\lambda/2$ and $\lambda/4$ birefringence retardations between orthogonal linear polarization bases. Similarly, astigmatic transformation allows for closely analogous control over states of paraxial Gaussian mode, enabling the adiabatic connection of all feasible states on the same HLG modal sphere. More specifically, the transformation, when accompanied by a $\lambda/4$ astigmatic retardation, is capable of converting HG modes located on the equator into modes situated on other regions of the sphere (such as LG or HLG modes) and vice versa; while the transformation with a $\lambda/2$ astigmatic retardation serves to rotate HG modes or invert the helicity of other modes on the sphere.

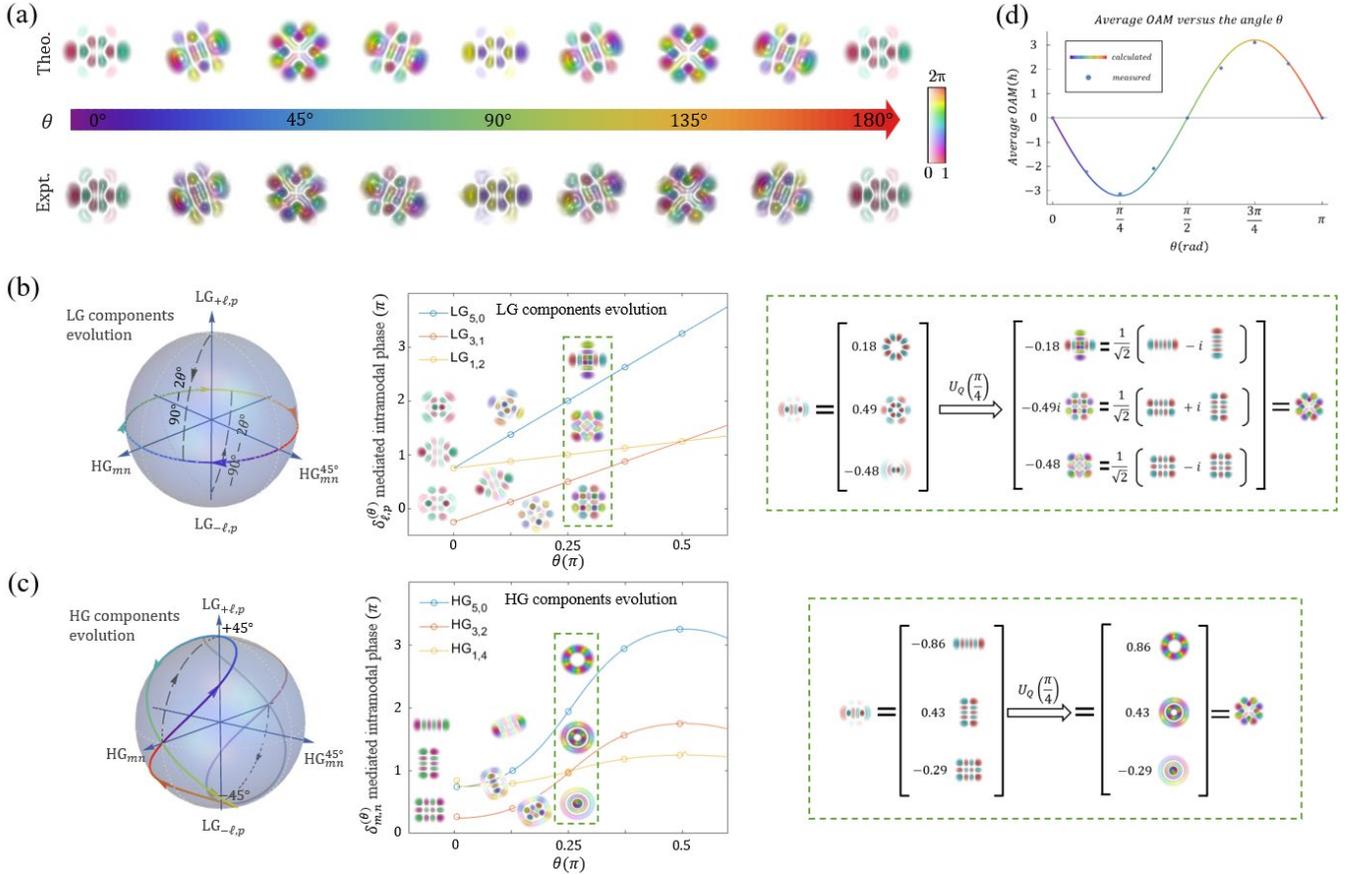

**Figure 2**. (a) Pattern evolution of the $IG_{5,3}$ mode under a single unitary transformation with a $\lambda/4$ retardation, the beamforming mechanism explained from LG (b) and HG (c) representations, and average OAM of the transformed beam with $\theta$.

### A. Pattern variations in single λ/4 transformation

The following discussion focuses on the modal conversion of complex structured Gaussian modes through a single unitary transformation with a $\lambda/4$ retardation. In Figure 2(a), the simulated and observed pattern evolution of the example mode $IG_{5,3}$ as a function of the astigmatic retarder's slow-axis angle ($\theta$) is presented. The results demonstrate that the spatial complex amplitude undergoes a continuous variation as over the slow-axis angle varies from 0 to $\pi/4$. Subsequently, a mirror inversion evolution occurs. Importantly, all patterns revealed through the unitary transformation exhibit a propagation invariant beam structure, indicating they belong to the set of PWE eigen solutions as well. The beamforming mechanism that governs the evolution of patterns can be effectively elucidated by analyzing the $\lambda/4$ transformation on the superposition state presented in Eq. (4). Specifically, the spatial spectrum transformations after the unitary transformation, denoted as $U_Q(\theta)$, can be mathematically described by employing LG or HG representations, given by

$$\sum c_{\ell,p} \mathrm{LG}_{\pm\ell,p}(\boldsymbol{r}) \xrightarrow{U_Q(\theta)} \sum c_{\ell,p} e^{i\delta_{\ell,p}^{(\theta)}} \mathrm{HG}_{m,n}^{\pm 45°-\theta°}(\boldsymbol{r}) \quad (5)$$

or

$$\sum c_{m,n} \mathrm{HG}_{m,n}(\boldsymbol{r}) \xrightarrow{U_Q(\theta)} \sum c_{m,n} e^{i\delta_{m,n}^{(\theta)}} \mathrm{HLG}_{m,n}^{(\theta)}(\boldsymbol{r}) \quad (6)$$

where $\mathrm{HG}_{m,n}^{\pm 45°-\theta°}(\boldsymbol{r})$ and $\mathrm{HLG}_{m,n}^{(\theta)}$ denote $\pm 45°-\theta°$ rotated $\mathrm{HG}_{m,n}^{\square}$ and intermediate HLG components, respectively, $\delta_{\ell,p}^{(\theta)} = \ell\theta$ and $\delta_{m,n}^{(\theta)} = |m-n|\tan^{-1}(\tan^2\theta)$ are geometric phase acquired by corresponding components [26]. The Eqs. (5) and (6) above clearly demonstrate that the modal components and their intramodal phase, i.e., $\mathrm{Arg}(c_{\ell,p})\delta_{\ell,p}^{(\theta)}$ or $\mathrm{Arg}(c_{m,n})\delta_{m,n}^{(\theta)}$, both vary with the slow-axis angle $\theta$, which directly contributes to the corresponding variations observed in the global pattern.

Figure 2(b) and (c) depict the beamforming mechanism, illustrating the spatial-complex amplitudes of modal components and their relative phase as they continuously vary with the angle $\theta$. The utilization of either HG or LG as modal components, i.e., the two representations, offers distinct viewpoints on the beamforming mechanism. However, it is noteworthy that despite these disparities, their conclusions are identical. Specifically, the LG viewpoint, illustrated in Fig. 2(b), demonstrates the direct transport of all modal components from the two polars to the equator along corresponding lines of $\pm 90° - 2\theta°$ longitude. As a result, three initial pairs of LG conjugate, i.e., $\mathrm{LG}_{\pm\ell,p} = 1/\sqrt{2}(\mathrm{LG}_{+\ell,p} + \mathrm{LG}_{-\ell,p})$, are transformed into rotating orthogonal HG superpositions, i.e., $1/\sqrt{2}(\mathrm{HG}_{m,n}^{+45°-\theta°} + e^{i2\theta}\mathrm{HG}_{m,n}^{-45°-\theta°})$, as exemplified in the right green dashed box. In contrast, the HG viewpoint offers an alternative analysis, depicted in Fig. 2(c), wherein the increase of $\theta$ leads to a gradual conversion of three initial HG components into LG modes along corresponding geodesics on their respective modal spheres.

Notably, analysis from both viewpoints predicts that the newly revealed patterns in Fig. 2(a) carry net OAM. Specifically, the amount of average OAM per photon ($\bar{\ell}\hbar$) increases with $\theta$ and reaches its maximum at $\theta = 45°$ or $135°$, where the initial parity modal components $\mathrm{HG}_{m,n}$ and $\mathrm{LG}_{\pm\ell,p}$ are transformed into the corresponding maximum helical modes $\mathrm{LG}_{\ell,p}$ and *helical* HG modes $1/\sqrt{2}(\mathrm{HG}_{m,n}^{\square} \pm i\mathrm{HG}_{n,m}^{\square})$, respectively. Figure 2(d) presents the calculated and measured average OAM of the transformed IG modes as a function of the angle $\theta$, and the calculations obtained from both viewpoints are identical. The measured results, represented by dots on the theoretical curve, were obtained through projecting observed spatial-complex amplitudes shown in Fig. 2(a) on LG bases [28,29]. In summary, the observed pattern evolution of the example IG mode in a single $\lambda/4$ transformation can be attributed to the fact that all of its modal components undergo an astigmatic conversion and acquire an OAM-dependent geometric phase.

### B. Pattern revivals in cyclic transformation

We further discuss the pattern evolution of complex structured Gaussian modes in cyclic transformations, where the concluding state of each HLG component remains identical to the initials state, except for the addition of an overall geometric phase. The magnitude of the phase shift is determined by both the topological charge of polar LG conjugates and the geodesic loop on the sphere, given by

$$\phi_g(\ell, \Omega) = (|\ell| + 1)\Omega/2, \quad (7)$$

where $\Omega$ is the solid angle encompassed by the geodesic loop drawn by the cyclic transformation. Figure 3 illustrates two viable approaches for achieving the cyclic transformation which involve cascading unitary transformations using combinations of quarter-quarter-half (QQH) and quarter-half-quarter (QHQ) retarders, with LG and HG modes serving as the starting points on the geodesic loop, respectively. The pattern evolutions of the superposition state in a cyclic transformation, specifically the example $IG_{5,3}$ mode, can be described as

$$\sum c_{m,n} \mathrm{HG}_{m,n}(\boldsymbol{r}) \xrightarrow{U_{\mathrm{QQH}}(\Omega)} \sum c_{m,n} e^{i\phi_g(\ell,\Omega)} \mathrm{HG}_{m,n}(\boldsymbol{r}) \quad (8)$$

$$\sum c_{\ell,p} \mathrm{LG}_{\ell,p}(\boldsymbol{r}) \xrightarrow{U_{\mathrm{QHQ}}(\Omega)} \sum c_{\ell,p} e^{i\phi_g(\ell,\Omega)} \mathrm{LG}_{\ell,p}(\boldsymbol{r}) \quad (9)$$

when considering the two approaches. The equations above illustrate that while the magnitude of each modal component remains constant in both transformation approaches, their intramodal phases are influenced by the geometric phase outlined in Eq. (7). Consequently, as the solid angle $\Omega$ increases, the superposition state undergoes a continuous variation in pattern and experiences structural revivals, as well as inverse image, at specific values.

presented in Figs. 4(a) and (b), which depict the simulated and observed pattern evolution of the example mode during the transformation along the two approaches, respectively. The phase evolution of LG and HG components retrieved from the observation, obtained from projection of spatial complex amplitudes, confirms the theoretical result in Fig. 3. In addition to proving the inference regarding pattern revival, pattern variations observed in the two approaches exhibit distinct behavior. Of particular interest is that the QHQ approach exhibits an evolution characterized by an image rotation of the global pattern, with the pattern revival occurring as the spin angle reaches $2\pi$.

This phenomenon can be adequately explained using the transformation outlined in Eq. (9). More specifically, the geodesic loop of a QHQ transformation on a modal sphere originates from and ultimately returns to the LG polar. Consequently, each LG component of the IG mode (as well as other complex structured light) undergoes a phase shift depending on its OAM, thereby inducing the rotation of the global pattern [17]. Similarly, it is well known that the $\lambda/2$ transformation *always* demonstrates image rotation [30], as LG modes are either transported to the opposite polar in a single operation or undergo cyclic transport through double $\lambda/2$ transformations. This implies that the physics underlying image rotation lies in the fact that OAM states and their representation can be considered as eigen modes and the parameter space of the image rotation operation. All the aforementioned principles are applicable to any complex modal structured light that can be expressed as a superposition of HLG modes.

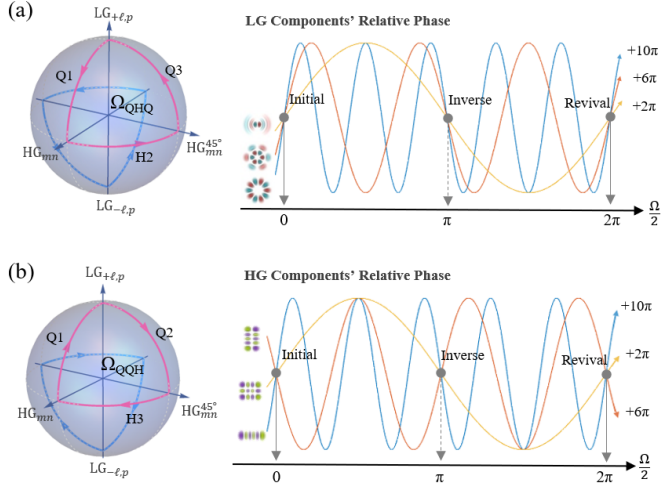

**Figure 3**. Two approaches (a) QHQ and (b) QQH used to achieve the cyclic transformation. Right-side curves illustrate variations of LG and HG components' relative phase with the increase of $\Omega$, in which gray arrows points positions of initial, inverse imaging, and revival states.

Pattern revivals occur when all the phase shifts are equal to multiples of $2\pi$. For the example mode $IG_{5,3}$, as shown in Fig. 3(a), its first pattern revival is expected to occur at $\Omega = 4\pi$, i.e., path $\lambda/2$ transformation (i.e., $H_2$ or $H_3$ on spheres) twice surrounding the equator, where the geometric phase acquired by its three LG or HG components is $2\pi$, $6\pi$, and $10\pi$, respectively. This prediction is confirmed by the results

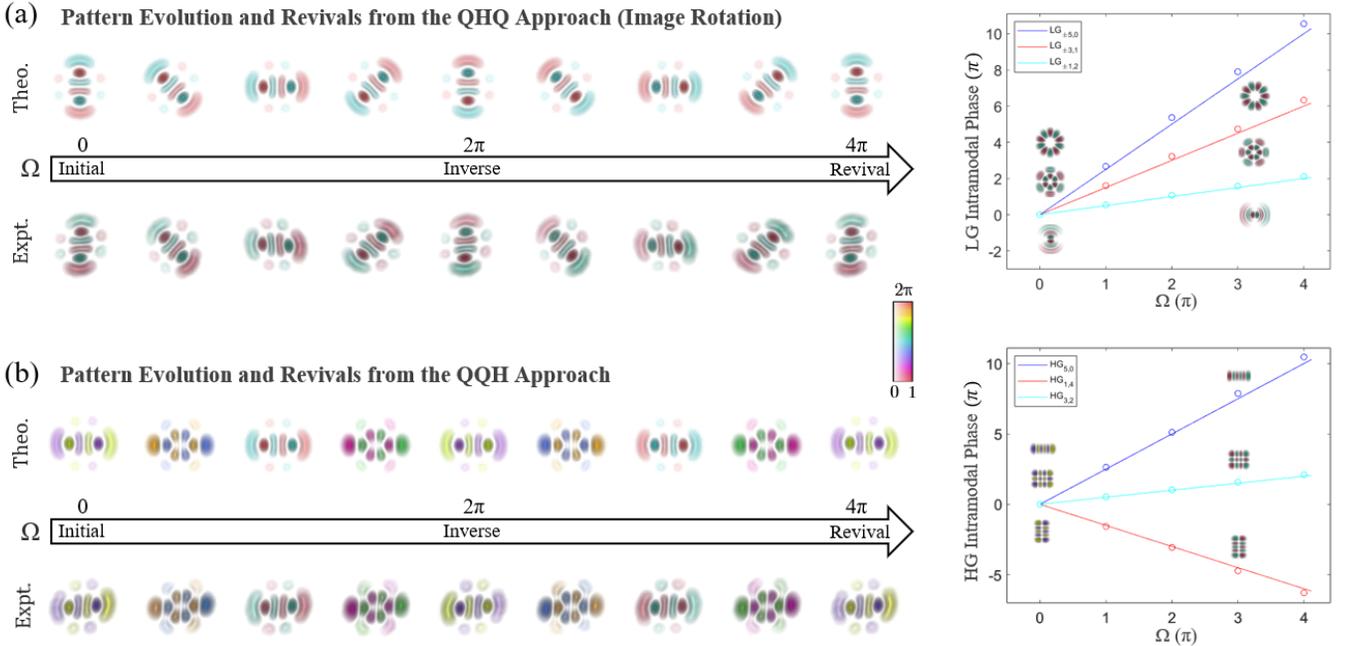

**Figure 4**. Pattern evolution and revivals of the $IG_{5,3}$ mode in cyclic transformation upon (a) QHQ and (b) QQG approaches, and associated intramodal phase evolution mediated by nonsynchronous geometric phase. The empty circles embed on corresponding color lines are measured intramodal phases obtained from the observed complex amplitudes.

## IV. Conclusion

The aforementioned demonstrations provide evidence that, first, the unitary modal evolution of a complex structured Gaussian beam can be effectively elucidated by considering the astigmatic transformation of its HLG component, as well as their intramodal phase variation with the geometric phase of the transformation. Second, it is observed that in the case of cyclic transformations, each HLG component remains constant while acquiring a non-synchronized geometric phase as described in Eq. (7). Consequently, it is reasonable to anticipated that the global pattern will undergo structural revivals when all the phase shifts are equal to multiples of $2\pi$. Furthermore, the implementation of cyclic transformations in the LG or other OAM representations lead to the image rotation of the global pattern.

Analogous to the temporal distortion observed in dispersive propagation of mode-locked pulses, the variations and revivals of the global pattern observed in this study can be interpreted as the spatial distortion of 'mode-locked' patterns, which essentially are paraxial eigen modes or their superpositions with the identical order, induced by spatial-mode dispersion. Considering this perspective, it can be anticipated that in the context of ultrafast scenarios, beyond the pattern evolution, the pattern would gradually lose its visibility during the transformation, because the temporal overlap between modal components decreases as well, i.e., becomes a mixed state [31,32]. In addition to 'mode-locked' patterns, above beamforming mechanism also applies to paraxial non-eigen modes and, particularly, the pattern evolution in cyclic transformations could be used to explore states of optical high-dimensional topology [13–15].

In summary, we have presented a comprehensive study on the unitary modal evolution of complex structured Gaussian beams using IG modes as an illustrative example, which are superposition states comprised of HLG modes on distinct modal spheres. The Talbot-like modal evolution of these complex structured Gaussian beams, exhibiting in pattern variations and revivals, has been systematically analyzed and experimentally examined. The essence of the pattern evolution lies in the intramodal phase dynamics of HLG components, which is non-synchronously mediated by the higher-order geometric phase of OAM degrees of freedom. Our results outline a generic modal conversion theory of structured Gaussian beams through astigmatic unitary transformation. The beamforming mechanism behind the pattern variations and revivals offers a new beamforming approach for spatial modal shaping and control, and has great potential in a wide variety of applications with structured light.

## Appendix: Experimental Details

Figure 5 depicts the experimental setup used to validate the theoretical analysis, which was comprised completely by transmission elements. The initial state $IG_{5,3}$ was generated by utilizing a liquid-crystal planar element (LC-1) based on geometric phase [33], which was illuminated by a $TEM_{00}$ beam operating at a wavelength of 780 nm. Subsequently, the state was directed into the apparatus for the unitary transformation, and the configuration presented here is tailored for the QQH cyclic transformation. More specifically, to achieve cascading $\lambda/4$ transformations, planar retarders fabricated with liquid-crystal geometric phase (LCQ) was used, see Ref. 24 for details. The $\lambda/2$ control was achieved with a Dove prime. The output beams' spatial complex amplitudes were then characterized using a Stokes tomography system [28].

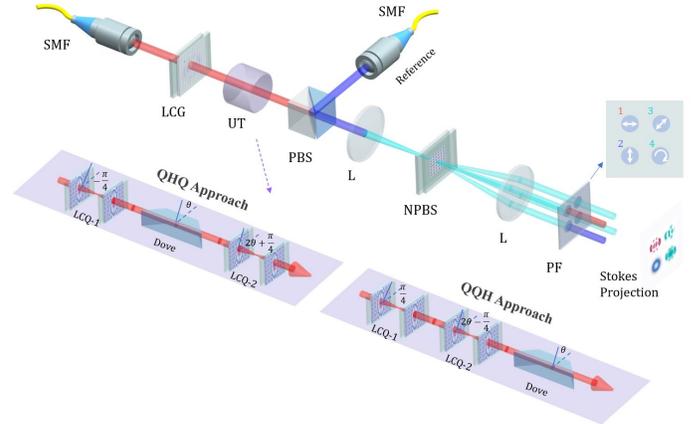

**Figure 5**. Schematic representation of the experimental setup, where key components include a polarizing beam splitter (PBS), non-polarizing beam splitter (NPBS), polarization filter (PF), single-mode-fiber collimators (SMF), lenses (L), and the unitary transformation module (UT) including two astigmatic $\lambda/4$ retarders (LCQ) and a Dove prime.

## Acknowledgments

Z.-H. Z acknowledges financial support from the National Natural Science Foundation of China (Grant Nos. 62075050 and 11934013).